\documentclass[10pt,a4paper]{article}
\usepackage[utf8]{inputenc}
\usepackage{amsmath}
\usepackage{amsfonts}
\usepackage{amssymb}
\usepackage{xfrac}
\usepackage{bm}
\usepackage{upgreek}
\usepackage{graphicx}
\usepackage[all]{nowidow}
\usepackage[left=2.5cm,right=2.5cm,top=2.5cm,bottom=2.5cm]{geometry}
\usepackage{cite}
\usepackage{dirtytalk}
\usepackage{float}
\usepackage[doublespacing]{setspace}
\usepackage{authblk}

\begin{document}

\title{Weak localisation enhanced ultrathin scattering media}

\author[1]{R. C. R. Pompe\thanks{contributed equally to this work}}

\author[2]{D. T. Meiers$^*$}

\author[1]{W. Pfeiffer}

\author[2,3]{G. von Freymann}

\affil[1]{Department of Physics, Bielefeld University, 33615 Bielefeld, Germany}
\affil[2]{Physics Department and Research Center OPTIMAS, Technische Universit\"at Kaiserslautern, 67663 Kaiserslautern, Germany}
\affil[3]{Fraunhofer Institute for Industrial Mathematics ITWM, 67663 Kaiserslautern, Germany}

\maketitle
\newpage

\textbf{The brilliant white appearance of ultrathin scattering media with low refractive index contrast and the underlying radiative transport phenomena fascinate scientists for more than a decade. Examples of such systems are the scales of beetles of the genus \textit{Cyphochilus}\cite{Vukusic07,Luke10}, photonic network structures \cite{Utel19} or disordered Bragg stacks (DBS) \cite{Meiers18,Rothammer21}.
While previous studies relate the highly efficient scattering in the scales to the anisotropy of the intra-scale network and diffusive light transport \cite{Burresi14, Cortese15, Wilts18, Burg19, Jacucci19, Lee20, Lee21}, the coherent radiation propagation dynamics remained unaccounted for. Here, we identify different coherent light transport regimes using time and spatially resolved coherent light scattering spectroscopy. At least 20\% of the collected scattered light originates from weakly localised random photonic modes, in contrast to solely diffusive light transport assumed to date \cite{Burresi14, Cortese15, Lee20, Lee21}. The identification of this significant role of weak localisation in ultrathin brilliant scattering media establishes a new design paradigm for efficient scattering optical materials.}\\

\begin{figure}
    \centering
    \includegraphics[width=\textwidth]{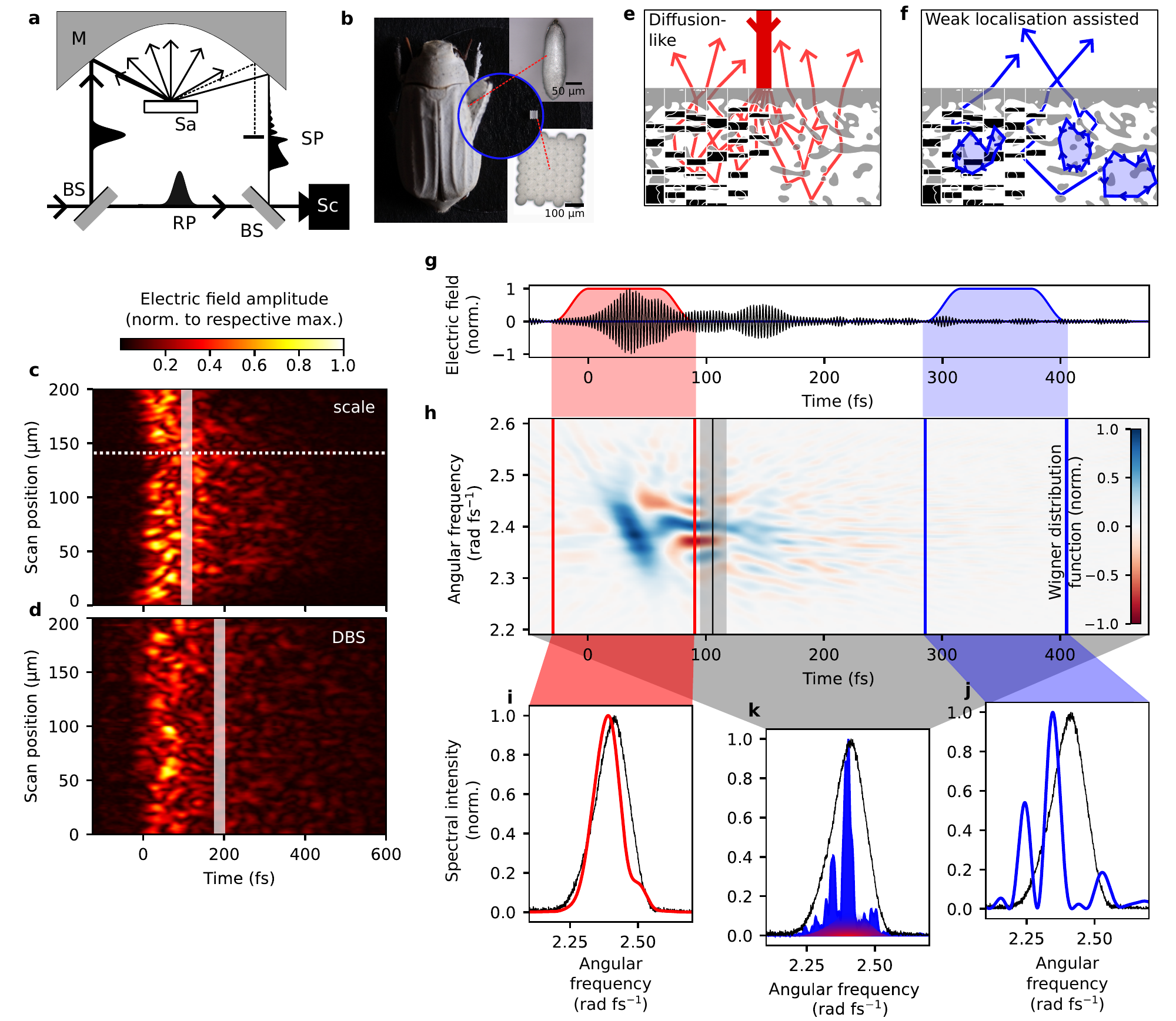}
    \caption{\textbf{Microscopic and ultrafast time-resolved spectroscopy of light  scattered from  \textit{Cyphochilus} scales and microfabricated DBS structures.} 
    \textbf{a}, Scheme of the spectral interference setup (see explanations in the text and in Methods).
    \textbf{b}, Photograph of \textit{Cyphochilus} (left) and disordered Bragg stacks (DBS, centre closeup) with light microscope images of a single beetle scale (right top) and DBS (bottom) as insets.
    \textbf{c,d}, Spatially resolved time domain amplitude of light scattered from a \textit{Cyphochilus} scale (\textbf{c}) and DBS (\textbf{d}). The transition threshold between diffusive regime and resonance radiation as identified in \textbf{h} are indicated (vertical translucent bar).
    \textbf{e,f}, Scheme illustrating how incoming light is scattered in the initial diffusion-like regime (\textbf{e}) and later via weakly localised photonic modes indicated by closed pathways (\textbf{f}). The grey structure is a cross section of a \textit{Cyphochilus} scale (taken from Wilts \textit{et al.} \cite{Wilts18}). The black overlay on the left side shows the disordered Bragg stacks.
    \textbf{g}, Scattered electric field at a single scan position (white dashed line in \textbf{c}) with indication of the short time Fourier transform windows used in \textbf{i} (red) and \textbf{j} (blue).
    \textbf{h}, Wigner distribution function of the scattered field shown in \textbf{g}. At $105\pm10$ fs (black line) the dominating light transport regime changes from diffusion-like to weak localisation assisted.
    \textbf{i,j,k}, Fourier spectra of the early time window (\textbf{i}) (-50 to 50 fs, red in \textbf{g} and \textbf{h}), the later time window (\textbf{j}) (250 to 350 fs window, blue in \textbf{g} and \textbf{h}) and the total measured time window (\textbf{k}).
    \label{fig:fig1}
    }
\end{figure}

In strongly scattering media the description of light propagation as ballistic transport breaks down and is commonly replaced by diffusive radiation transport that explains well the observed optical characteristics in numerous applications \cite{Schittny16,Lorenzo12}. Diffusive radiation transport neglects the coherent propagation of scattered fields and hence does not account for interference phenomena in disordered media, which are known to occur for example when weak localisation gives rise to coherent back scattering \cite{Kaveh86} or random lasing in disordered active media \cite{Wiersma08}.
For increased scattering strength coherent back scattering occurs, when two counter-propagating scattering light paths in the medium, i.e. the illuminating light and collinear back scattered light, interfere constructively and giving rise to a peak in the back scattered intensity, as it was, e.g., reported for \textit{Cyphochilus} scales \cite{Jacucci19}. However, modelling of the brilliant white appearance of \textit{Cyphochilus} scales still completely relies on diffusive propagation \cite{Burresi14, Cortese15, Lee20, Lee21} and thus coherent effects are neglected. This could hamper tailoring disordered photonic media since an unambiguously identified scattering mechanism is the basis for nanostructure design for optimised performance. Using ultrafast time-resolved light scattering spectromicroscopy \cite{Differt15, Aeschlimann15} we here identify the coherent light scattering mechanisms for \textit{Cyphochilus} scales and disordered Bragg stacks and show that weak localisation in leaky photonic modes significantly contributes to the brilliant whiteness of these scatterers.

The identification of coherent scattering is significantly facilitated if the number of interfering pathways is kept small. For example, laser speckles are most pronounced when only a small area of the scatterer is illuminated. However, if the detector integrates over sufficiently many different interfering pathways, the speckles disappear. In this case with exception of the coherent back scattering peak, the scattering behaviour is often well explained by diffusive radiation transport theory, although the underlying transport is coherent. To reduce the number of interfering pathways the present investigation relies, both, on focused illumination and collection of scattered light from a small sample volume. Furthermore, coherent propagation adds a well-defined phase to the scattered fields and thus reconstruction of the temporal evolution of the scattered electric field provides additional information on the scattering mechanism.

To systematically study the impact of coherent transport on the whiteness of the \textit{Cyphochilus'} scales, we use the setup shown in Fig. \ref{fig:fig1}a to perform ultrafast time-resolved light scattering spectromicroscopy on a single scale\cite{Differt15,Aeschlimann15}. The observations are confirmed for DBS fabricated via direct laser writing (see Supplementary Information) shown in Fig. \ref{fig:fig1}b. The DBS mimic the beetle scales, reproduce their known optical properties \cite{Meiers18} and allow for realistic scattering light simulations based on finite-difference time-domain (FDTD) Maxwell solvers and Monte Carlo (MC) diffusive light transport simulations.

To achieve the spatial resolution necessary to observe only few interfering scattering pathways, a parabolic mirror (Fig. \ref{fig:fig1}a, M) focuses a pulsed Ti:sapphire laser beam down to a $\lesssim 3$ µm spot on the surface of the sample (Sa) and collects the scattered light under an angle of $\sim 24$° relative to the specular direction. To filter for intra-scale scattering, i.e. multiple scattered light components, a cross-polarisation configuration is used. The illuminated position is scanned by moving the sample using a piezo stage. Spectral interference \cite{Lepetit95} between the scattered light pulse (SP) and a reference pulse (RP) allows for the time reconstruction of the field of the scattered light (see Methods). The amplitude of the measured electric field (cf. Fig. \ref{fig:fig1}c and d) shows for both samples essentially the same dynamics, i.e. spatially varying exponential decay modulated by distinct beating, indicating interference taking place. As discussed below two different propagation regimes can be identified in the scattered light signals. Initially diffusion-like transport (Fig. \ref{fig:fig1}e) dominates, whereas for longer times radiation leaking from weakly localised photonic modes formed by randomly closed scattering pathways (Fig. \ref{fig:fig1}f) prevails, which gives rise to the observed beating behaviour.

To identify the different propagation regimes we analyse the coherent scattering signal (cf. Fig. \ref{fig:fig1}g) in time and frequency domain by means of the Wigner distribution function (WDF, see Methods) \cite{Mecklenbraeuker97}, exemplarily shown in Fig. \ref{fig:fig1}h for the \textit{Cyphochilus} scale. For early times broadband features are present, which reproduce the excitation spectrum when evaluating the short time Fourier transform (cf. Fig. \ref{fig:fig1}i). At about $105\pm10$ fs there is a qualitative change in the spectral content of the WDF, i.e. broad spectral features are replaced by fine modulations. This time matches closely to the pulse round trip time (see Methods), i.e. the time a pulse needs to travel back and forth through the layer assuming a homogeneous, effective medium with an effective refractive index, as it is commonly done in diffusion approximation. The spectral modulations stem from multiple sharp resonances, which become better visible in the short time Fourier transform for later times (cf. Fig. \ref{fig:fig1}j). The power spectrum illustrates that the signal now contains spectral peaks independent of the original excitation spectrum, whereas the scattered light in the initial diffusion-like phase exhibits no significant modulation. The spectrum for the full measured signal, shown in Fig. \ref{fig:fig1}k, exhibits spectral peaks on top of a broadband background and thus reflects the spectral characteristics of both transport regimes.

\begin{figure}
    \begin{center}
    \includegraphics[width=\textwidth]{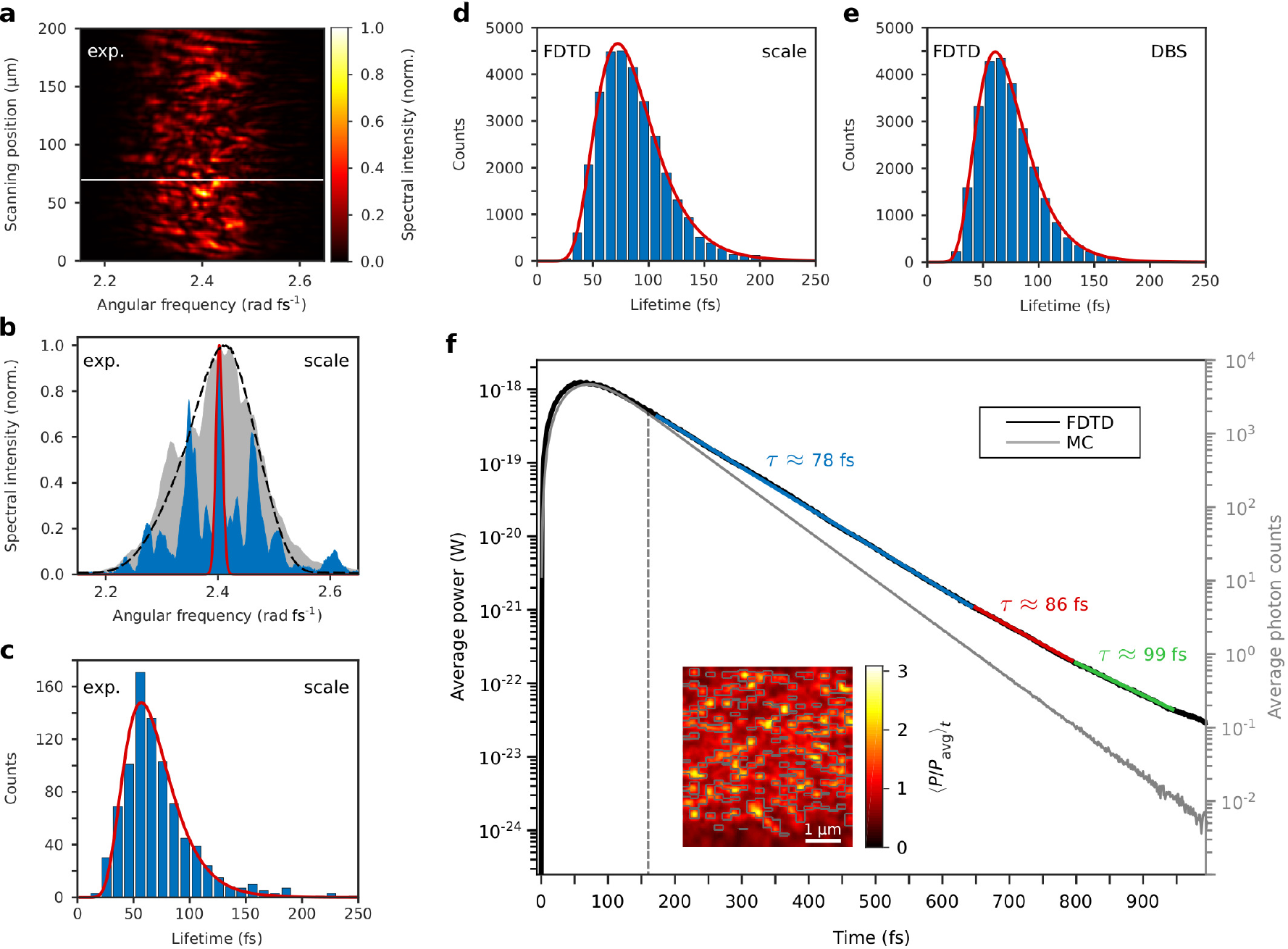}
    \caption{\label{fig:fig2} \textbf{Lifetime distribution for weakly localised photonic modes.} 
    \textbf{a}, Spatially resolved light scattering spectra of \textit{Cyphochilus} scale. 
    \textbf{b}, Spectral intensity (in blue) for the position indicated by the white line in \textbf{a}. The excitation spectrum and the incoherent mean spectral intensity over the entire scan is shown as dashed line and grey shaded area, respectively. Distinct peaks are identified (exemplified by red curve) and used to estimate the corresponding photonic mode lifetimes.
    \textbf{c}, Photonic mode lifetime distribution derived from the scan displayed in \textbf{a}. 
    \textbf{d, e}, Lifetime distributions obtained from FDTD simulations of the intra-scale structure \cite{Wilts18} (\textbf{d}) and the DBS model (\textbf{e}). 
    \textbf{f}, Transient average power in the monitor plane perpendicular to the surface sectioning the DBS model (cf. Fig. \ref{fig:Details_FDTD}a) derived from FDTD simulation (black curve) and average photon counts in the same plane calculated by Monte Carlo simulation (grey curve). Both ordinates span the same orders of magnitudes, making the slopes directly comparable. The non-exponential decay of the FDTD results is indicated by coloured exponential slopes with different lifetimes $\tau$. The vertical dashed line indicates the point in time where both curves start to differ. Inset: The time averaged local power enhancement in a snippet of the FDTD monitor plane averaged over the time span indicated by the blue line (170-650 fs).}
    \end{center}
\end{figure}

While the short time Fourier transforms (Fig. \ref{fig:fig1}i,j) allow identifying the contribution of the different light transport mechanisms over time, this spectral analysis of resonances lacks of resolution due to the short time windows. To unambiguously identify the weak localisation assisted scattering the probability distribution of the resonance lifetimes is investigated applying full time Fourier transformations. Fig. \ref{fig:fig2}a reveals that the scattered light spectra possess multiple peaks with varying centre frequency and width as function of the spatial coordinate. In the incoherent mean of the spectra over the whole scan (Fig. \ref{fig:fig2}b, grey shaded area) these narrow spectral peaks average out and reproduce the excitation spectrum (Fig. \ref{fig:fig2}b, dashed line), macroscopically resulting in the white appearance. Based on peak fitting (Fig. \ref{fig:fig2}b, red curve) we derive the spectral widths of the peaks, which yield a lower limit for the underlying resonance lifetimes. The distribution of these lifetimes is displayed in Fig. \ref{fig:fig2}c and follows a log-normal distribution (red curve), deviating from a normal distribution for longer lifetimes as expected when localisation effects occur \cite{Pinheiro08}. The tail towards long lifetimes is associated with the rare occurrence of increasingly localised modes, i.e. cases where scattering pathways close inside the structure instead of coupling to loss channels \cite{Mascheck12}.

This identification of weak localisation assisted light scattering is further supported by FDTD simulations based on the known microstructure of the \textit{Cyphochilus} scale \cite{Wilts18} (model data provided by courtesy of B. Wilts) and the DBS. As exemplified in Fig. \ref{fig:Details_FDTD}b and c the local spectra recorded inside the structures also exhibit sharp resonances. Statistical analysis of these resonances yields the lifetime distributions shown in Fig. \ref{fig:fig2}d and e, which are in excellent accordance with the experimental results. Hence, we conclude that the spectral resonances experimentally observed in the scattered light indeed originate from weakly localised photonic modes occurring in the same way inside the beetle structure and DBS. The corresponding spectral features give rise to the observed beating behaviour in scattered light spectromicroscopy (Fig. \ref{fig:fig1} c,d).

To further investigate the light propagation inside the structure the spatio-temporal evolution of the local power (in FDTD simulations) and the photon counts (in MC simulations) are recorded on a monitor plane sectioning the DBS perpendicular to the surface (cf. Fig. \ref{fig:Details_FDTD}a). To avoid artefacts from the lateral periodic boundary conditions (see Methods) a sufficiently large lateral simulation domain of $20\times20$ µm² is used. This ensures that any potential spectral contribution from this periodicity lies far outside the considered spectral range. In contrast to the rather complex beetle intra-scale structure the DBS consist of simple building blocks and thus is used for further simulations to keep the computation time manageable. 

The FDTD simulations (Fig. \ref{fig:fig2}f, black curve) reveal a non-exponential decay with lifetimes $\tau$ ranging from about 80 fs up to roughly 100 fs. This directly reflects the lifetime distribution (Fig. \ref{fig:fig2}e) possessing a mean value around 80 fs, implying that for longer times the longer living photonic modes dominate the decay. In contrast the MC simulations (Fig. \ref{fig:fig2}f, grey curve) show a mono-exponential decay with a decay constant of 65 fs (cf. Fig. \ref{fig:comparison_ls}a), failing to match both the simulated and measured lifetime distributions. Nevertheless, it is possible to find a set of parameters such that the MC simulations reproduce for the same layer thickness the properties of the DBS obtained by FDTD simulations, i.e. reflectance, transport mean free path and initial shape of the curve. Hence, we conclude that the initial coherent transport inside the structure can be approximated as diffusive transport emphasising that there is a diffusion-like scattering regime despite interference effects may occur. However, beyond about 170 fs modelling as diffusive transport breaks down and the curve obtained by MC simulation starts to deviate from the FDTD results. Assuming propagation in an effective medium approach (as done for the experiment) yields a pulse round trip time of 160 fs for the 100 fs long pulses applied in the simulations (see Methods). This coincides well with the time at which FDTD and MC simulations deviate indicating that the pulse round trip time is indeed a suitable estimation for the upper limit of the time domain in which diffusion-like photon transport dominates. For longer times the trapping in weakly localised photonic modes takes over, which is only captured in the fully coherent FDTD simulations.

The FDTD simulations provide means to directly visualise the weakly localised photonic modes inside the DBS structure (inset in Fig. \ref{fig:fig2}f). The time averaged local power enhancement normalised to the average power (see Supplementary Information) exhibits distinct, spatially localised hotspots with an up to three times enhanced local power. These hotspots are associated with antinodes of weakly localised random photonic modes (as depicted schematically in Fig. \ref{fig:fig1}f) which give rise to the experimentally observed distinct peaks in the spectra (cf. Fig. \ref{fig:fig1}j). As expected incoherent diffusive photon propagation in MC simulations do not exhibit any hotspots but an almost constant photon count enhancement across the monitor plane (cf. Fig. \ref{fig:comparison_ls}c). 

\begin{figure}[t]
    \centering
    \includegraphics[width=\textwidth]{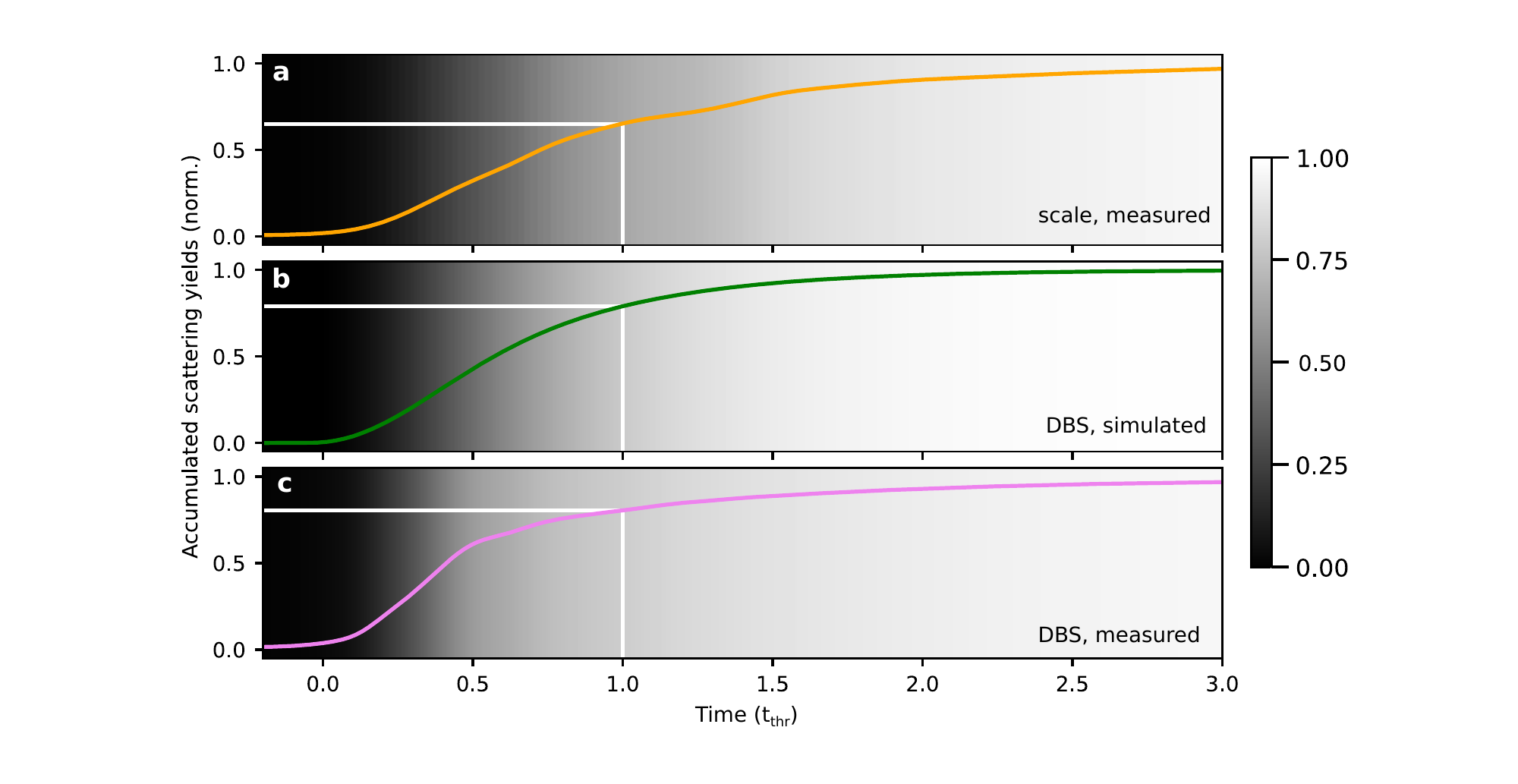}
    \caption{\textbf{Spatially averaged time-dependent accumulated scattering yields.}
    The square modulus of the time-resolved scattering fields are averaged over the recorded positions. This incoherent intensity signal is integrated over time to yield the time-resolved accumulated scattering yield. The background shading at $t_{\text{thr}}$ indicates the loss of whiteness if weak localisation assisted scattering would be absent.
    \textbf{a}, Accumulated scattering yield experimentally measured for the \textit{Cyphochilus} scale. The white vertical line corresponds to a threshold time of $t_{\text{thr}}$=105 fs, as indicated in Fig. \ref{fig:fig1}b,h, from which one weak localisation scattering dominates. The scattering yield from weak localisation is $35\%$ (white horizontal line).
    \textbf{b}, Accumulated scattering yield for the simulated DBS, with a threshold time of $t_{\text{thr}}$=160 fs, as indicated in Fig. \ref{fig:fig2}f. The scattering yield from weak localisation is $21\%$.
    \textbf{c}, Accumulated scattering yield experimentally measured for the fabricated DBS, with a threshold time of $t_{\text{thr}}$=190 fs (see Supplementary Information), as indicated in Fig. \ref{fig:fig1}c. The scattering yield from weak localisation is $20\%$.
    }
    \label{fig:fig3}
\end{figure}

Summarising the observations and model simulations we conclude that the scattering yield is dominated by photon leakage from weakly localised photonic modes after an initial scattering time window, which can be roughly estimated as the pulse round trip time in the ultrathin scattering layer treated in an effective medium approach. Such modes have previously been identified for systems that exhibit random lasing with coherent feedback \cite{Cao00, Wiersma08}, but were not yet identified to significantly contribute to the brilliant whiteness of ultrathin scattering media. As shown in Fig. \ref{fig:fig3} scattering via weakly localised photonic modes is responsible for at least about $20\%$ of the total scattering and thus is relevant when the scattering efficiency of ultrathin disordered photonic media are concerned. As indicated in the background shadings of Fig. \ref{fig:fig3} the scales and the DBS would appear rather greyish and not brilliant white, if scattering via leakage from weakly localised photonic modes would be missing.

In conclusion, we have experimentally shown that the light transport in scattering, brilliant white structures is dominated initially by a diffusion-like transport which is surpassed by scattering via leakage from weakly localised photonic modes after roughly the pulse round trip time in the ultrathin scattering layer. Leakage from weakly localised modes accounts for at least 20\% of the scattered light, underlining their significance for the brilliant whiteness of the ultrathin scattering media. This identification of the coherent weak localisation assisted scattering mechanisms based on time-resolved scattered light spectromicroscopy could serve, both conceptionally and methodologically, to gain a better understanding of the transport regimes in disordered materials and their time dynamics. This is  e.g. relevant in imaging through turbid media for bioimaging applications or random lasing action in disordered gain media \cite{Das03,Li17,Hohmann21}. Furthermore, the here demonstrated weak localisation feature of the biomimetic DBS relying on a distorted Bragg reflector design provides a blueprint for tailoring nanostructures to particularly support random photonic resonances which can enhance light-matter interaction and therefore may find applications as materials for efficient solar energy harvesting \cite{Differt15,Zhou18,Loh21} or sensor applications, where resonance enhanced absorption is employed to improve sensitivity \cite{Kassa20}.

\section*{Methods}

\textbf{Experimental setup.}
The light source is a mode-locked Ti:sapphire laser (Femtosource Scientific, Femtolasers Produktions GmbH, Austria) with a centre wavelength of $\lambda_0=780$ nm and spectral full width half maximum (FWHM) $\Delta \lambda = 47$ nm, filtered in s-polarisation relative to the sample. To achieve microscopic resolution the beam is focused onto the sample by a parabolic mirror (custom fabricate, Jenoptik, Germany). The sample is moved via a piezo stage (M-664.164, Physik Instrumente (PI) GmbH \& Co. KG, Germany) in the focal plane to scan the excitation and light collection position. The parabolic mirror horizontally separates the incoming beam, the specular reflection and the scattered light under different angles, allowing to select the measured scattering angle via a blocker aperture. To ensure that only light that was scattered multiple times is measured, the scattered light is measured in cross polarisation with a spectrometer (USB 2000, Ocean Optics Inc., USA).

\textbf{Phase reconstruction.}
The time resolution is achieved by phase reconstruction via spectral interference of the scattered light with a reference pulse. Therefore the incoming pulse is separated into sample and reference path. The reference path is delayed relative to the sample pulse and rotated into the measured p-polarisation.
The resulting interference spectrum $|E_{\text{s}}(\omega)+E_{\text{r}}(\omega)|^2=|E_{\text{s}}(\omega)|^2+|E_{\text{r}}(\omega)|^2+ E_{\text{s}}(\omega)E_{\text{r}}^*(\omega) \cos(\Delta \varphi(\omega))$ contains the phase difference $\Delta \varphi$ between the two beams. Via Fourier filtering of the interference spectrum  and after correcting for the phase imbalance of the interferometer the phase effect of the sample alone can be reconstructed (see Supplementary Information). Since the phase difference is measured no phase optimisation of the probing pulse is necessary.

\textbf{Wigner distribution function and Short Time Fourier Transform.}
The Wigner distribution function is defined as $W(t,\omega)=\int^{\infty}_{-\infty} E(t-t'/2)E^*(t+t'/2)\exp{(-i\omega t')}\text{d}t'$, where $E$ and $E^*$ are the complex electric field and its complex conjugate respectively. The WDF yields the highest time-frequency resolution possible. On the other hand it is not a linear transform, resulting in cross-terms modulating the the WDF. To help with the interpretation the spectral power of the short time Fourier transform (STFT), given by $|S(\tau,\omega)|^2=|\int^{-\infty}_{\infty} w(t',\tau,\Delta t, t_r)E(t')\exp{(-i\omega t')} \text{d}t'|^2$, where $w(t,\tau,\Delta t, t_r)$ is a Tukey window function \cite{Harris78} centred at time $\tau$, is used, which as linear transform produces no cross-terms. For the STFT the spectral resolution is limited by the window width $\Delta t = 120$ fs. The window rising time is $t_r=30$ fs.

\textbf{Calculation of the pulse round trip time.}
For a single photon travelling back and forth through an effective medium with thickness $l_{\text{s}}$ the effective round trip time is given by $t_{\text{eff}} = {2l_{\text{s}}}/{v_{\text{eff}}}$. The speed of light inside the medium is calculated via $v_{\text{eff}} = c_0/n_{\text{eff}}$ where the effective refractive index $n_{\text{eff}}$ is computed using the Maxwell-Garnett mixing rule \cite{Ruppin00}. To obtain the limit when all photons within the pulse length have propagated back and forth through the effective medium, i.e. the pulse round trip time $t_{\text{prt}}$, the pulse length has to be added to the effective round trip time of a single photon. This ensures that also the `last' photon within the pulse length has reached the top of the medium again. 
The scale and the simulated DBS structure possess a filling fraction of $f_{\text{scale}} = 31\%$ \cite{Burg19,Lee20} and $f_{\text{DBS}} = 27\%$, respectively and the refractive index of chitin $n_{\text{chitin}} = 1.55$ \cite{Leertouwer11} is used in both cases. Applying these values in the Maxwell-Garnett mixing rule yields $n_\text{eff, scale} = 1.15$ for the scale as well as $n_\text{eff, DBS} = 1.13$ for the DBS. Evaluating the effective round trip time with a sample thickness of $l_{\text{s, scale}} = 10$ µm \cite{Burg19} and $l_{\text{s, DBS}} = 7.9$ µm results in $t_{\text{eff, scale}} = 77$ fs for the scale and $t_{\text{eff, DBS}} = 60$ fs for the DBS, respectively. The pulse length is defined as the time span between the pulse front and the point in the pulse tail where the intensity dropped to $I_{\text{p}}/e^2$ with the peak intensity of the pulse $I_{\text{p}}$. In the experiment the pulse front is set at the point where the intensity first reaches $I_{\text{p}}/e^2$ yielding a pulse length of $t_{\text{pulse, exp}} = 29$ fs. In the simulation the definite pulse front as emitted by the source is used, resulting in a pulse length of $t_{\text{pulse, sim}} = 100$ fs. Thus, pulse round trip times of $t_{\text{prt, scale}} = 106$ fs and $t_{\text{prt, DBS}} = 160$ fs are obtained for the scale and DBS, respectively.

\textbf{Extraction of lifetimes from spectral peaks.}
We estimate the intensity lifetimes of the resonances by $\tau_l = 1/\Delta \omega$, where $\Delta \omega$ is the spectral intensity FWHM of the peak \cite{Mascheck12}. To measure the spectral widths of a peak, it is fitted with a Gaussian (cf. Fig. \ref{fig:meth_peakfit}). Fitting the individual peaks ignores slope change by overlapping resonances, thus the resulting lifetimes are accordingly lower estimates. To identify individual peaks in the frequency-position plane of the line scans a 2D peak finding routine is used.

\textbf{Finite-difference time-domain simulations.}
The FDTD simulations were performed using the software Lumerical FDTD Solutions (Ansys Inc., USA). In all simulations a plane wave pulse impinges in the z-direction on the respective structure (cf. Fig. \ref{fig:Details_FDTD}a). In the z-direction we apply perfectly matched layers as boundary conditions. In the x- and y-direction we use periodic boundary conditions to eliminate unwanted absorption in lateral boundaries due to the finite size of the simulation.
For the calculation of the lifetime distribution we collect the spectra from roughly 3900 distinct point-shaped frequency monitors placed in the structure model provided by Wilts \textit{et al.} \cite{Wilts18} and the DBS structure (for model parameters see Ref. \cite{Meiers18}) respectively, both occupying a footprint of $7\times7$ µm² and a height of $7 - 8$ µm. For excitation we use a light pulse with a centre wavelength of 780 nm and collect wavelengths between 745 nm and 815 nm approximating the experimental conditions. 
The calculation of the time-dependent power distribution is done for a DBS model based on the same parameters but with a lateral footprint of about $20\times20$ µm². A time-domain monitor cross sectioning the structure in the x-z-plane is applied to record every 1.14 fs the poynting vector at every monitor grid point over a total simulation time of 1000 fs. A pulse length of 100 fs is used to obtain a spectral narrow band excitation with a centre wavelength of 780 nm and a FWHM of 14 nm. The zero time is set to the time when the pulse front enters the structure.

\textbf{Monte Carlo Simulation.}
Monte Carlo simulations are performed using a self-written Matlab code (The MathWorks Inc., USA) based on the well known algorithm presented in literature \cite{Schittny16,Wang95}. To match the FDTD simulation conditions no absorption inside the slab is applied and in lateral direction periodic boundary conditions are used. As light source (with about 6.8 billion photons) a plane wave is chosen possessing a temporal profile matching the temporal power profile of the impinging pulse in FDTD simulations. An appropriate monitor cross sectioning the slab is placed according to the FDTD setup.
The lateral width of the slab is 12 µm, the height and effective refractive index are equal to the values given above for the simulated DBS model. The applied transport mean free path of $l_{\text{t}} = 3$ µm is equal to the one obtained by FDTD simulations (see Supplementary Information). A scattering mean free path of $l_{\text{s}} = 1$ µm is selected reproducing the FDTD results for short times closely (cf. Fig \ref{fig:comparison_ls}). The anisotropy factor $g$ is defined via $l_{\text{t}} = l_{\text{s}}/(1-g)$ \cite{Burresi14} and hence determined by the choice of $l_{\text{t}}$ and $l_{\text{s}}$.

\section*{Acknowledgements}

We gratefully acknowledge financial support from the German Research Foundation DFG within the priority program "Tailored Disorder -  A science- and engineering-based approach to materials design for advanced photonic applications" (SPP 1839). We thank B. D. Wilts for supplying us with a 3D computer tomography model of the beetle scales' inner structure. We thank the team of the Nano Structuring Centre (NSC) at the Technische Universität Kaiserslautern for their support with focused ion beam milling and scanning electron microscopy.
\\




\end{document}


\begin{center}
    {\Large \textbf{Weak localisation enhanced ultrathin scattering media}}
    
    {\Large Supplementary Information}
\end{center}

\medskip

\begin{center}
    {\large R. C. R. Pompe, D. T. Meiers, W. Pfeiffer, and G. von Freymann}
\end{center}

\medskip

\section{Experimental setup}
\begin{figure}[h]
\begin{center}
\includegraphics[width=.9\textwidth]{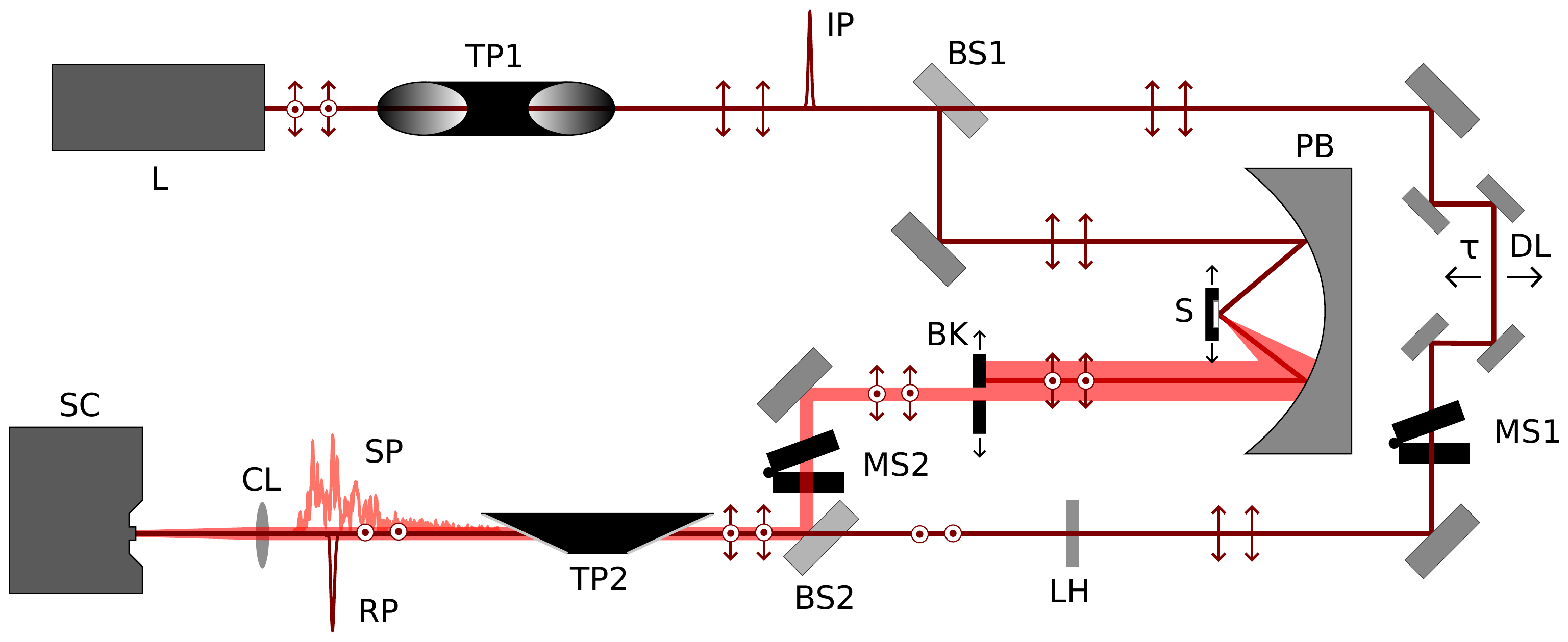}
\caption{\label{fig:meth_setup} \textbf{Scheme of the experimental setup.} The red arrows indicate polarisation, the black spatial translation. L: Laser; TP1, TP2: Thin film polariser; BS1, BS2: Beam splitter; PB: Parabolic mirror; S: Sample; BK: Reflected light blocker; DL: Delay stage; LH: $\lambda/2$ wave plate; MS1, MS2 : Mechanical shutter; CL: Collimating lens; SC: Spectrometer; IP: Incoming pulse; RP: Reference pulse; SP: Scattered light pulse. }
\end{center}
\end{figure}

\begin{figure}[h]
\begin{center}
\includegraphics[scale=.6]{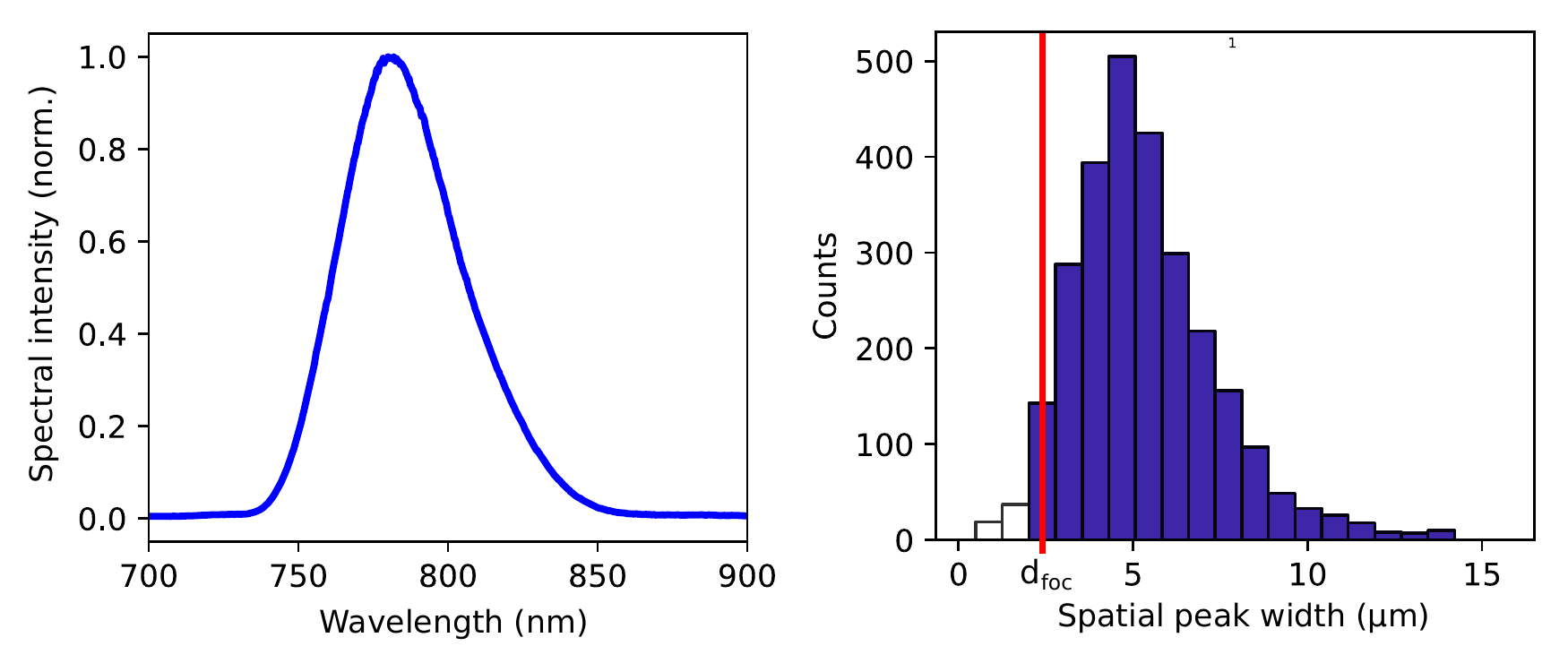}
\caption{\label{fig:meth_specandspot} \textbf{Excitation spectrum and spatial peak width of scattered spectra.}
\textbf{a}, Excitation spectrum, normalised to the maximum. 
\textbf{b}, Distribution of spatial widths of spectral peaks, used for spot size estimation. The red line indicates the theoretical diffraction limited spot diameter $d_{\text{foc}}$. Combined values for respectively three $200$ µm line scans on \textit{Cyphochilus} scale and DBS.}
\end{center}
\end{figure}

As excitation source we use a  mode-locked  Ti:sapphire laser (Femtosource Scientific, Femtolasers Produktions GmbH, Austria, Fig. \ref{fig:meth_setup}) L) with a centre frequency of $\lambda_0 = 780$ nm and a repetition rate of $78$ MHz, operated at a minimal spectral FWHM of $47$ nm (see Fig. \ref{fig:meth_specandspot}a) and filtered for s-polarisation by a thin film polariser (TP1). The incoming pulse (IP) is guided into an interferometer setup with a delay to adjust the interference spectrum (DL) in one arm and a custom made parabolic mirror (PB, Jenoptik, Germany), focusing the excitation beam onto the sample, in the other arm. The sample (S) is fixed on a piezo stage (M-664.164, Physik Instrumente (PI) GmbH \& Co. KG) to horizontally scan the excitation position. The parabolic mirror translates the scattering angle into a lateral displacement, allowing to choose the measured scattering angle with a blocker aperture (BK) of 3 mm in diameter. The scattering spectrum is measured in cross (p-)polarisation (TP2) to filter for surface scattering. Accordingly the reference polarisation is rotated by a $\lambda/2$ wave plate (LH). The interference spectrum resulting from the coherent superposition of sample and reference beam is recorded with a spectrometer (USB 2000, Ocean Optics Inc., USA).\\

To estimate the actual laser spot size on the surface of the sample the spatial widths of the spatially resolved spectra are determined by fitting (see sec. \ref{sec:lifetimeext}). The distribution shows a distinct drop at peak widths of $\sim 2.5 $µm (Fig. \ref{fig:meth_specandspot}b). With a diameter of the beam when hitting the parabolic mirror of $d_{\text{in}}=6$mm, the focal length of the parabolic mirror $f_{\text{pm}}=15$ mm and the wavelength at the lower limit of the FWHM of the excitation spectrum $\lambda_{\text{low}}=755$ nm result in a diffraction limited spot size of $d_{\text{foc}}=4 \lambda_{\text{low}} f_{\text{pm}}/\pi d_{\text{in}}= 2.4$ µm. Values below this threshold are considered to result from erroneous peak identifications and fits.

\section{Fabrication of the DBS structure}

\begin{figure}[h]
\begin{center}
\includegraphics[scale=1]{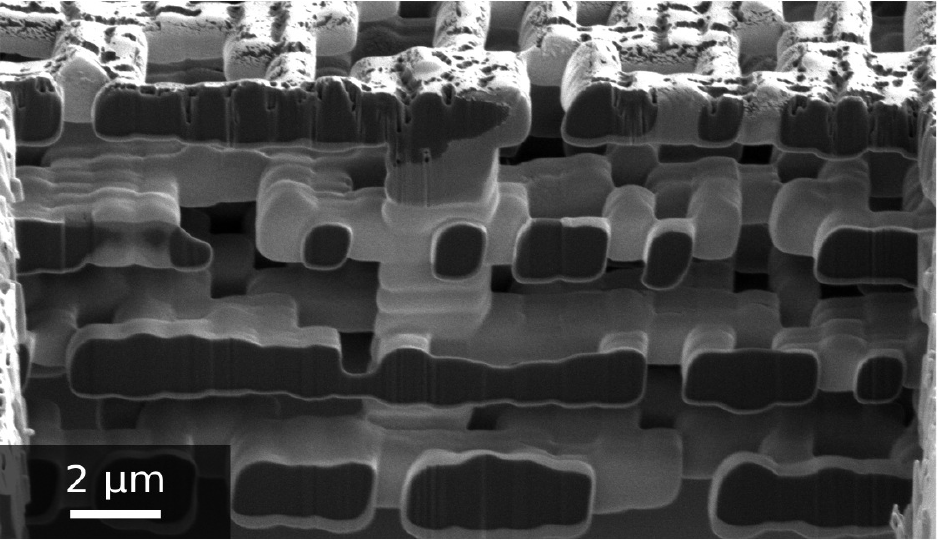}
\caption{\label{fig:DLW_DBS} \textbf{Cross section through the fabricated DBS structure.} SEM image of a direct laser written DBS's topmost four layers. The cross section is achieved using focused ion beam milling.}
\end{center}
\end{figure}

To fabricate the DBS structure via DLW we use a Photonic Professional GT (Nanoscribe GmbH, Germany) and the photoresist IP-L 780 (Nanoscribe GmbH). A drop of the photoresist is placed on top of a 170 µm thick coverslip (Gerhard Menzel GmbH, Germany) while the femtosecond pulsed laser beam (780 nm) is focused into the resist from the bottom side using immersion oil (Immersol 518F, Carl Zeiss AG, Germany) between the objective ($63\times$, NA = 1.4) and the glass substrate.
A mechanically stable DBS structure can only be produced for the model with constant centre-to-centre distance of consecutive layers as described in Ref. \cite{Meiers18_supps}. To obtain the separated layer of the DBS model first an open circular scaffold structure is written with a total diameter of 64 µm. Subsequently the top layer of the DBS structure is written inside the scaffold where the scaffold and the layer overlap 7 µm at the rim. Starting from the top layer four pillars (3.8 µm long, footprint $2\times2$ µm²) are added towards the substrate to achieve a constant spacing between the layers (cf. Fig. \ref{fig:DLW_DBS}). Following this procedure the next layers and pillars are written from top to bottom till the total amount of nine layers is reached. For every layer the laser power is varied over the writing position leading to different voxel sizes and thus to different thicknesses of the individual building blocks (cf. Fig. \ref{fig:DLW_DBS}). The footprint of a single building block is $1\times1$ µm², while the thickness mainly varies between 1 - 2 µm. To obtain a structure with an overall size of roughly $0.5\times0.5$ mm² 46 individual DBS structures are arranged in a hexagonal lattice as it can be seen in the inset of Fig. \ref{fig:fig1}b.

\section{Phase reconstruction}

The measured superposition of the scattered light field $|E_{\text{s}}|$ and the reference pulse field $|E_{\text{r}}|$ results in an interference spectrum $|E_{\text{s}}(\omega)+E_{\text{r}}(\omega)|^2=|E_{\text{s}}(\omega)|^2+|E_{\text{r}}(\omega)|^2+ E_{\text{s}}(\omega)E_{\text{r}}^*(\omega) \cos(\Delta \varphi(\omega))$ depending on the phase difference $\Delta \varphi(\omega)=\delta \varphi + \omega \tau$ between the two interferometer arms, where the linear phase $\omega \tau$ due to the delay in the reference arm is separated explicitly from the remaining phase difference $\delta \varphi$. Due to linearity the Fourier transform decomposes in components corresponding to $|E_{\text{s}}|^2+|E_{\text{r}}|^2$ and side bands $E_{\text{s}}E_{\text{r}}^*\exp(\pm(\delta \varphi + \omega \tau))$, centred around $\pm \tau$. For sufficiently high $\tau$  one side band can be isolated in Fourier space. Taking the argument of the back transformed isolated side band then yields $\delta \varphi= \varphi_\text{r} -(\varphi_\text{s}+ \varphi_\text{\text{sample}})$. Correcting with the phase difference of the interferometer arms, i.e. without sample, then yields the phase introduced by the sample $\varphi_\text{\text{sample}}$, independent of the total phase of illuminating and reference pulse.

\section{DBS signal time domain filtering}
\begin{figure}[h]
\begin{center}
\includegraphics[scale=.6]{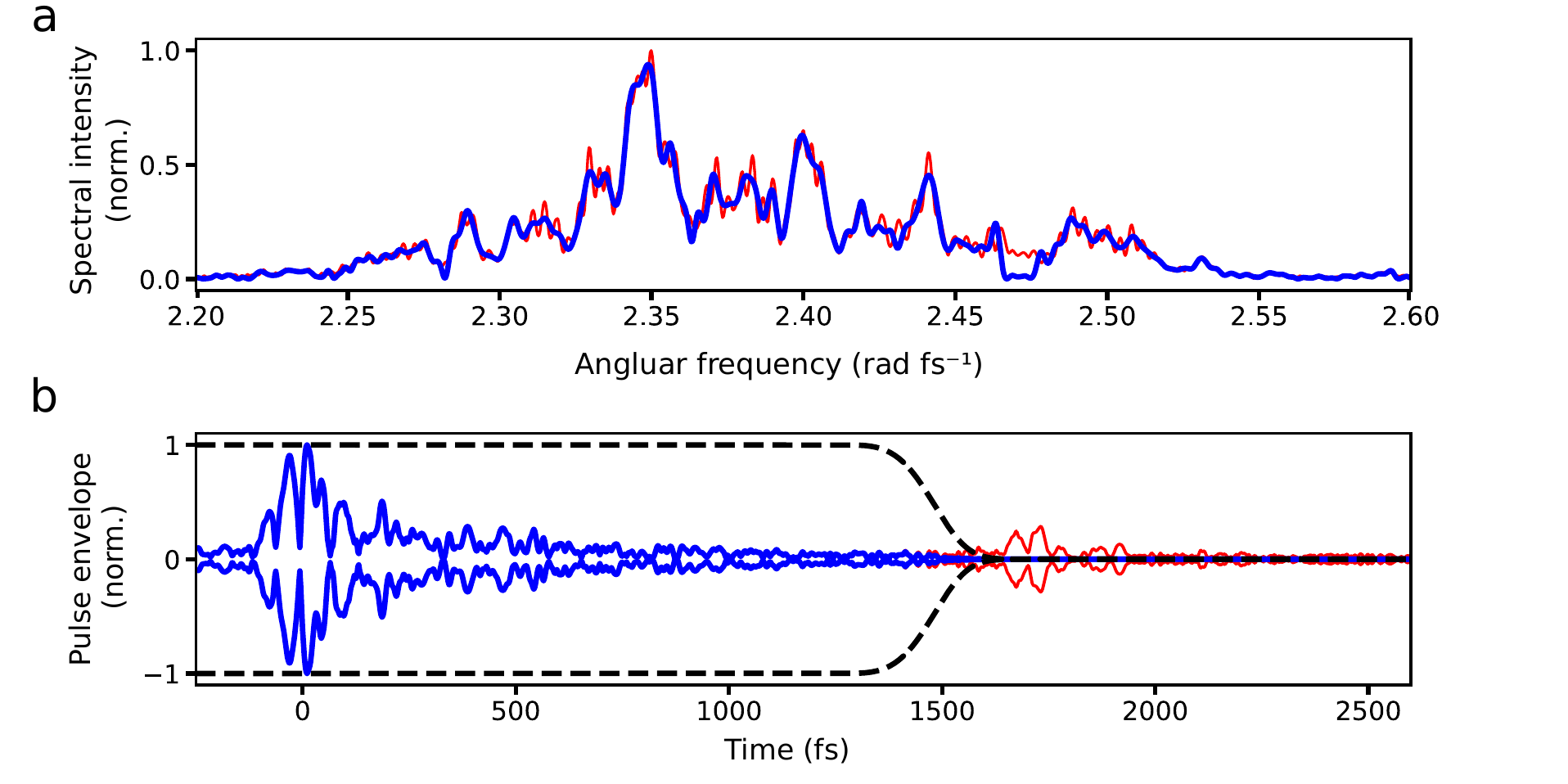}
\caption{\label{fig:meth_DBSfilterspec}
\textbf{Experimental time- and frequency-resolved scattered light signal for DBS}
\textbf{a}, Example of a scattered light spectrum of the DBS structure, including interference with the pulse reflected from the carrier slide (red), and time domain filtered, suppressing the second pulse (blue). 
\textbf{b}, Corresponding reconstructed time domain amplitude, including the reflected second pulse (red), and filtered (window, black dashed curve) for this unwanted contribution (blue curve).
}
\end{center}
\end{figure}
The DBS structure is written on a $d_{\text{slap}} = 170$  µm thin glass slap. To safely position the sample in the spectroscopic setup it is fixed on a standard microscope slide, introducing a mismatch in refractive indices. This results in an unwanted reflection over the whole scanned range, visible as fast modulation due to interference in the scattered light spectra  and a second pulse at $t_{\text{p2}}=1.8 \text{ ps} \approx 2 d_{\text{slap}}n_{\text{FS}}/c$ (fused silica, $ n_{\text{FS}}(\lambda=785\text{ nm}) =1.45$ \cite{Malitson65_supps}) in time domain (see Fig. \ref{fig:meth_DBSfilterspec}a and b), i.e. the ballistic portion of the pulse penetrating the DBS structure and travelling back and forth the thin slap. To obtain the spectral response of the DBS structure alone, the signals are filtered (black dashed line in Fig. \ref{fig:meth_DBSfilterspec}) in time domain and the intensity spectra of these time domain filtered signals are used for further processing.

\section{Extraction of life times}\label{sec:lifetimeext}

\begin{figure}[h]
\begin{center}
\includegraphics[scale=.8]{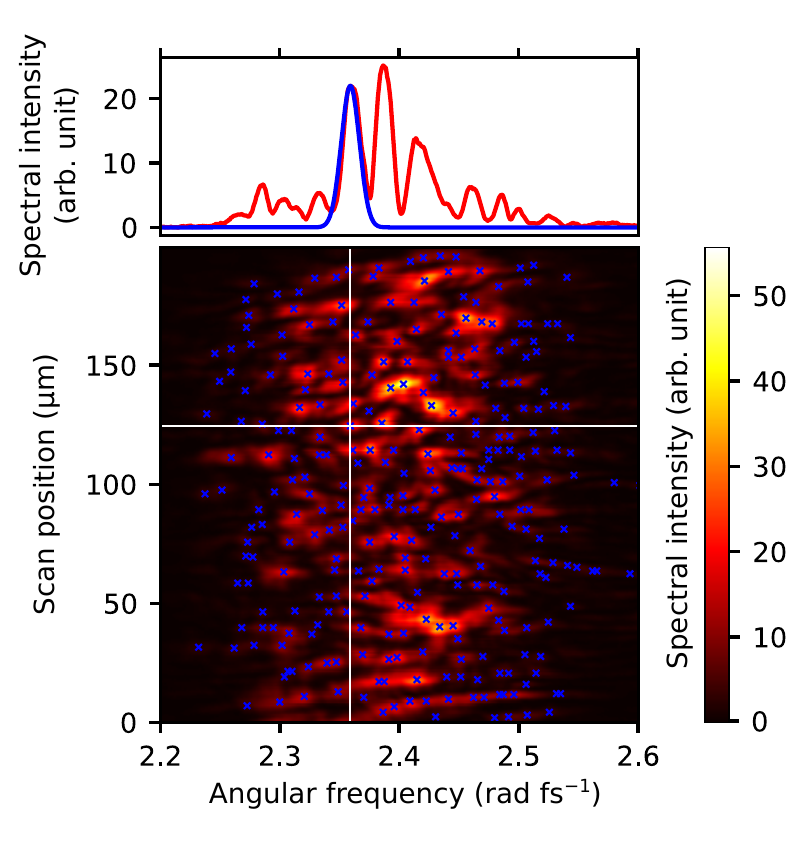}
\caption{\label{fig:meth_peakfit}
\textbf{Determination of spectral peaks and their widths.}
The blue crosses indicate the identified individual spectral peaks. The blue cross hair corresponds to the presented spectral intensity (top, red curve) and the peak that is fitted (top, blue curve).}
\end{center}
\end{figure}

We evaluate the intensity lifetimes of the resonances by $\tau_{\text{l}} = 1/\Delta \omega$, where $\Delta \omega$ is the spectral intensity FWHM of the peak. The spectral width of a peak is measured by fitting it with a Gaussian  (Fig. \ref{fig:meth_peakfit}) on half the range between the two neighbouring minima to account for peak shoulders etc. The Gaussians are used instead of Lorentzian, since the resonance line shapes are distorted due to overlap. The same procedure was applied on the spatial axis of the spectral line scans to extract the spatial peak widths, used to separate individual peaks along the spatial axis and to estimate the laser spot diameter on the surface of the samples. Fitting the individual peaks ignores peak slope change by overlap, thus the extracted life times represent lower limits of the actual resonance lifetimes.\\
The maxima in the frequency-position space were identified by a 2D peak finder. The minimal distance of the peaks in both directions was adjusted in iterative steps until the minimal separation was greater than the mean over the determined widths. Choosing the highest maximum within this frequency-position ellipses avoids considering a peak twice due to noise. For each sample three line scans, respectively displaced orthogonal to the scanning direction, were evaluated.\\

\section{Experimental results for the DBS}

\begin{figure}[H]
\begin{center}
\includegraphics[scale=.74]{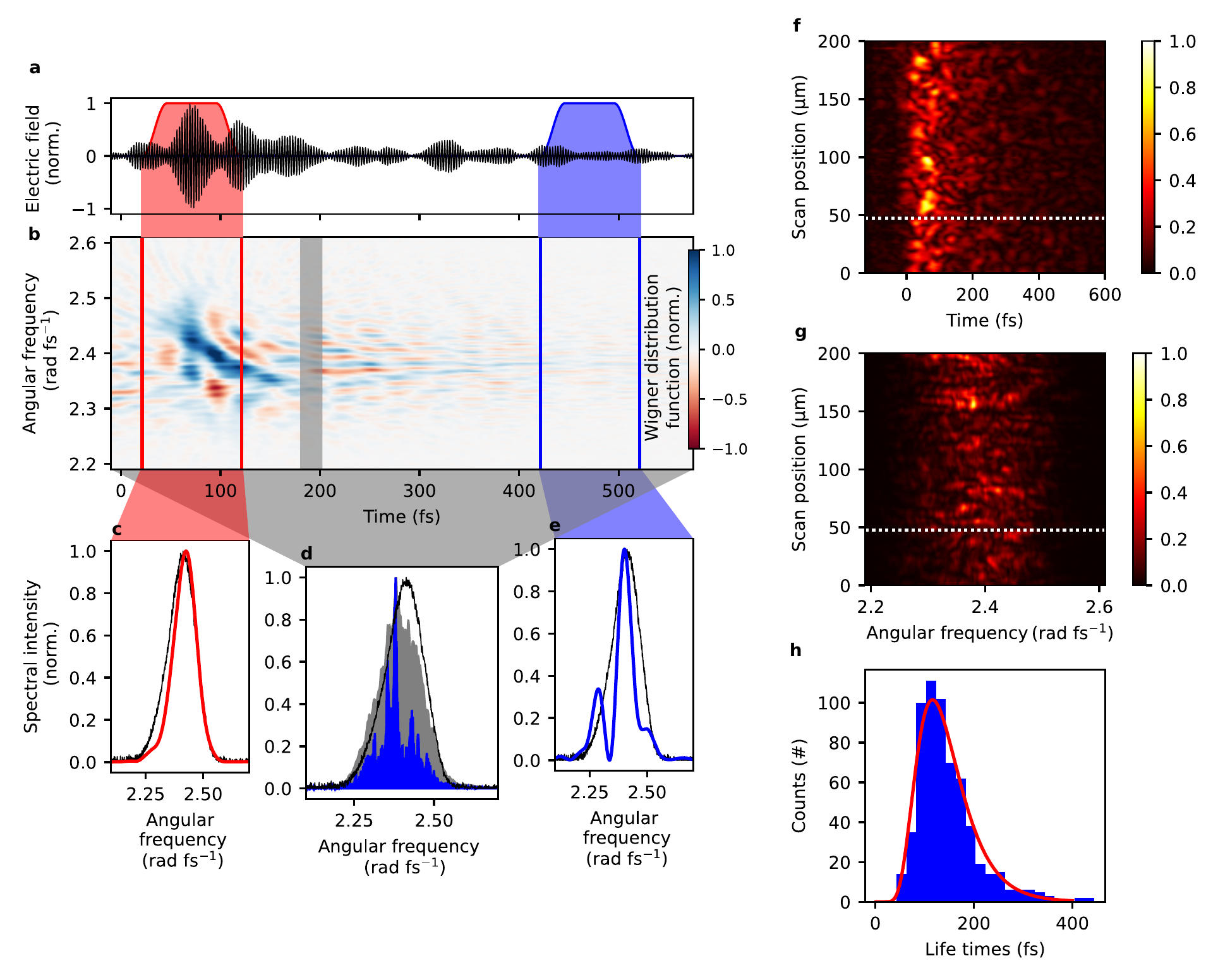}
\caption{\label{fig:realbragg_results}
    \textbf{Mircoscopic and ultrafast time-resolved spectroscopy of light scattered from microfabricated DBS}
    \textbf{a}, Normalised electric field of the scattered light at a single illumination position (white dotted line in \textbf{f}), with short time Fourier transform windows in red and blue.
    \textbf{b}, WDF of the signal shown in \textbf{a}. The translucent bar indicates the switch from diffusion-like to weak localisation assisted regime. The red and blue lines indicate the window widths used for the STFTs in \textbf{c} and \textbf{e} respectively.
    \textbf{c}, Short time Fourier transform power spectrum for a time window in the diffusion-like regime, with the excitation spectrum (black line).
    \textbf{d}, Power spectrum of the total measured time interval (blue shaded), with the excitation spectrum (black line) and the incoherent spatially averaged scattered spectrum (grey shaded). 
    \textbf{e}, Short time Fourier transform for a time window in the weakly localised regime.
    \textbf{f}, Spatially resolved time domain field amplitude of the scattered light. White dotted line corresponds to \textbf{a}-\textbf{e}.
    \textbf{g}, Spatially resolved power spectrum. White dotted line corresponds to \textbf{a}-\textbf{e}.
    \textbf{h}, Life time distribution, with log-normal fit (red curve). 
    }  
\end{center}
\end{figure}

Figure \ref{fig:realbragg_results} shows the results of performing the same experiment and analysis as presented in the main text for the \textit{Cyphochilus} scale on a fabricated DBS.
Figure \ref{fig:realbragg_results}a displays the scattered electric field at a specific scan position. Figure \ref{fig:realbragg_results}b shows the corresponding Wigner distribution function. Similar to the \textit{Cyphochilus} scale (Fig. \ref{fig:fig1}h) it starts with broadband features, separating into finer modulations with time. The threshold time when the initial broadband features which are attributed to diffusion-like transport, develops into the fine spectral features stemming from weak localisation assisted random photonic resonances is identified as $190\pm 10$ fs (grey translucent bar). To demonstrate that the initial portion of the signal is close to the excitation spectrum (black line) and not modulated by interference effects the STFT for the time interval 20 fs to 120 fs is calculated (Fig. \ref{fig:realbragg_results}c). The STFT for late times (Fig. \ref{fig:realbragg_results}e, on [$420$ fs, $520$ fs]) exhibits spectral peaks (with broadened spectral width due to the limited time window of the STFT), indicating leakage from weak localisation assisted photonic resonances at late times.
Note that the round trip time used for the beetle scale and the simulated, unscaled DBS to determine the end of the diffusion-like light propagation within the structures is not applicable for the fabricated DBS. The calculated round trip time in this case is $t_{\text{prt}}=326$ fs. Considering the Wigner distribution functions, we find that at this time the diffusion-like regime is over for more than 100 fs. Thus we use the qualitative changes in the WDF and STFT only to estimate the threshold time. This deviation between fabricated DBS and scale and simulated DBS is attributed to the fact that assuming an effective medium is questionable, since the internal structure is not on a sub-wavelength scale.
Fig \ref{fig:realbragg_results}f shows the scan of the time domain amplitude of the scattered light. Like for the beetle scale it is decaying exponentially, modulated by random beating varying with the scan position. This signal is attributed to the interference of multiple resonances.\\
The spatially resolved scattering spectra (Fig. \ref{fig:realbragg_results}g) show multiple sharp resonance peaks varying randomly in centre frequency and with position, in close similarity to the \textit{Cyphochilus} scale. Fitting the individual peaks to determine the lifetime distribution results in a log-normal distribution (Fig. \ref{fig:realbragg_results}h).

\section{Further FDTD simulations results}

\begin{figure}[H]
\begin{center}
\includegraphics[width=\textwidth]{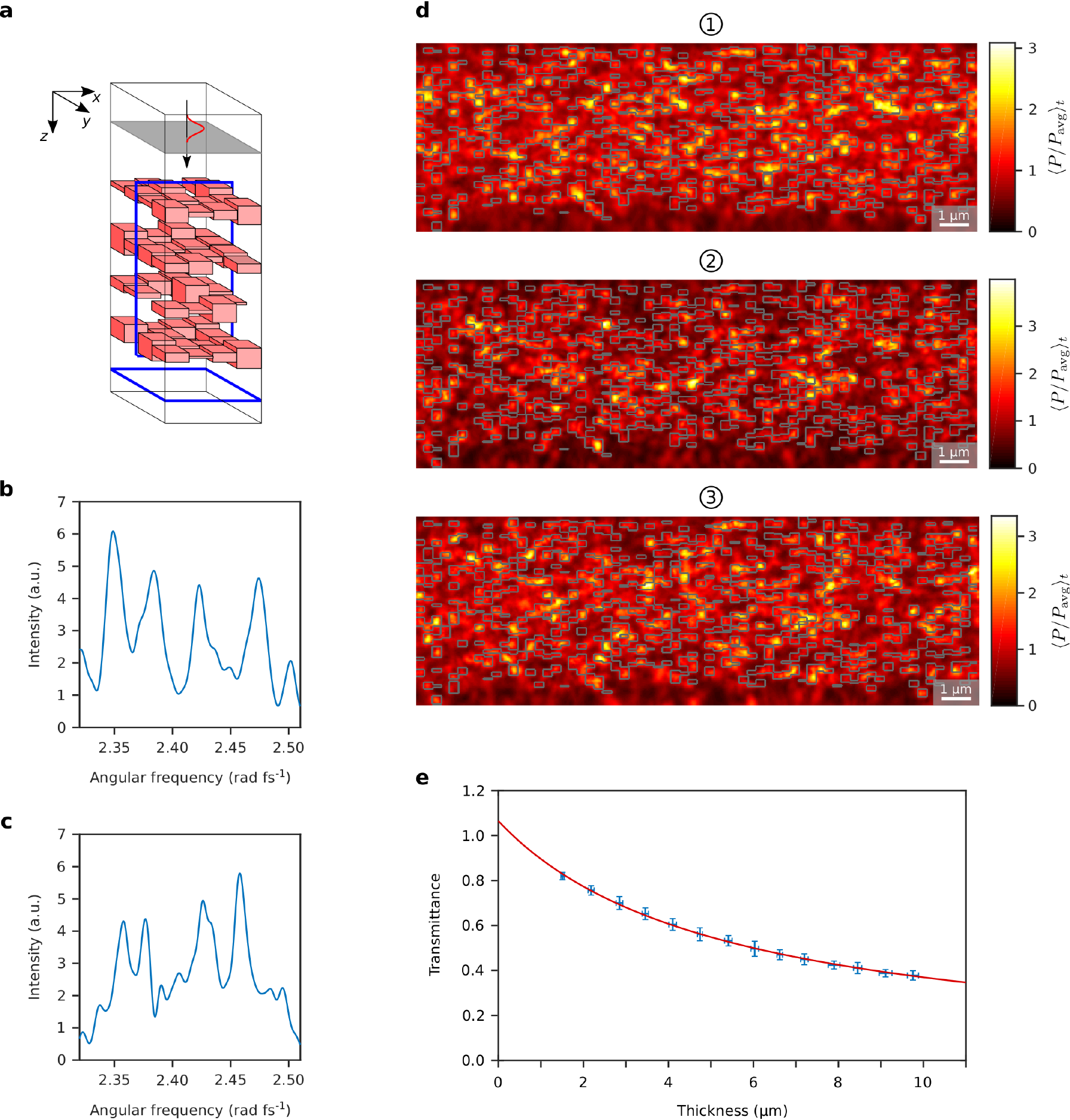}
\caption{\label{fig:Details_FDTD}
\textbf{Light transport regimes inside the simulated DBS.}
\textbf{a}, Sketch of the setup used for the simulation. A narrow band plane wave pulse impinges on the 3D DBS structure in the z-direction. During the propagation of the pulse the light power is time-resolved and spatially resolved collected by a x-z-monitor cross sectioning the structure as indicated by the lower horizontally oriented blue frame. To determine the transport mean free path the x-y-monitor denoted by the second blue frame is used to record the transmittance for different sample thicknesses.
\textbf{b,c}, Exemplary spectrum recorded with a point monitor inside the DBS (\textbf{b}) and intra-scale structure (\textbf{c}) in FDTD simulations. 
\textbf{d}, \raisebox{.5pt}{\textcircled{\raisebox{-.9pt} {1}}}, \raisebox{.5pt}{\textcircled{\raisebox{-.9pt} {2}}}, and \raisebox{.5pt}{\textcircled{\raisebox{-.9pt} {3}}} show the time averaged local power enhancement in the entire monitor plane for the time spans marked by the blue (170 - 650 fs), red (650 - 800 fs), and green (800 - 950 fs) line in Fig. \ref{fig:fig2}f, respectively.
\textbf{e}, DBS transmittance in dependence of the sample thickness fitted using the isotropic diffusion equation (red line). The error bars depict the standard deviation. 
}  
\end{center}
\end{figure}

To visualise temporally stable near field patterns, indicating standing waves in the nano-optical structures of the DBS (as shown in Fig. \ref{fig:fig2}f and Fig. \ref{fig:Details_FDTD}d), we record the time dependent power $P(x,z,t)$ over a cross section of the simulated structure (cf. Fig. \ref{fig:Details_FDTD}a, x-z-plane monitor). This quantity is normalised for every time step by the mean power $\bar{\rho}$, given by
\begin{align}
    \bar{\rho}(t)= \int_A P(x,z,t) \text{d}A/A,
\end{align}
to compensate for losses by radiation to the far field. This yields the momentary power enhancement
\begin{align}
    P_{\bar{\rho}}(x,z,t) = P(x,z,t)/\bar{\rho}(t).
\end{align}
Finally this quantity is averaged over the considered time interval $[t_1,t_2]$ to obtain  the time averaged power enhancement
\begin{align}
    \langle P/P_{\text{avg}} \rangle_t  = \langle P_{\bar{\rho}} \rangle_t = \int_{t_1}^{t_2} P_{\bar{\rho}}(x,z,t) \text{d}t/(t_2-t_1).
\end{align}

The transport mean free path is evaluated computing the transmittance for different number of layers in the DBS model and thus different structure thicknesses. Averaged over ten distinct realisations of the DBS model the transmittance $T$ is plotted against the according structure thickness $d$ as shown in Fig. \ref{fig:Details_FDTD}e. Fitting the data using isotropic diffusion theory $T = (l_{\text{t}} + h)/(d + 2h)$ \cite{Durian94} (red line) yields a transport mean free path of $l_{\text{t}} = (3.00 \pm 0.04)$ µm and an extrapolation length of $h = (2.65 \pm 0.09)$ µm.

\section{Determination of the scattering mean free path and photon counts enhancement}

\begin{figure}[h]
\begin{center}
\includegraphics[scale=1]{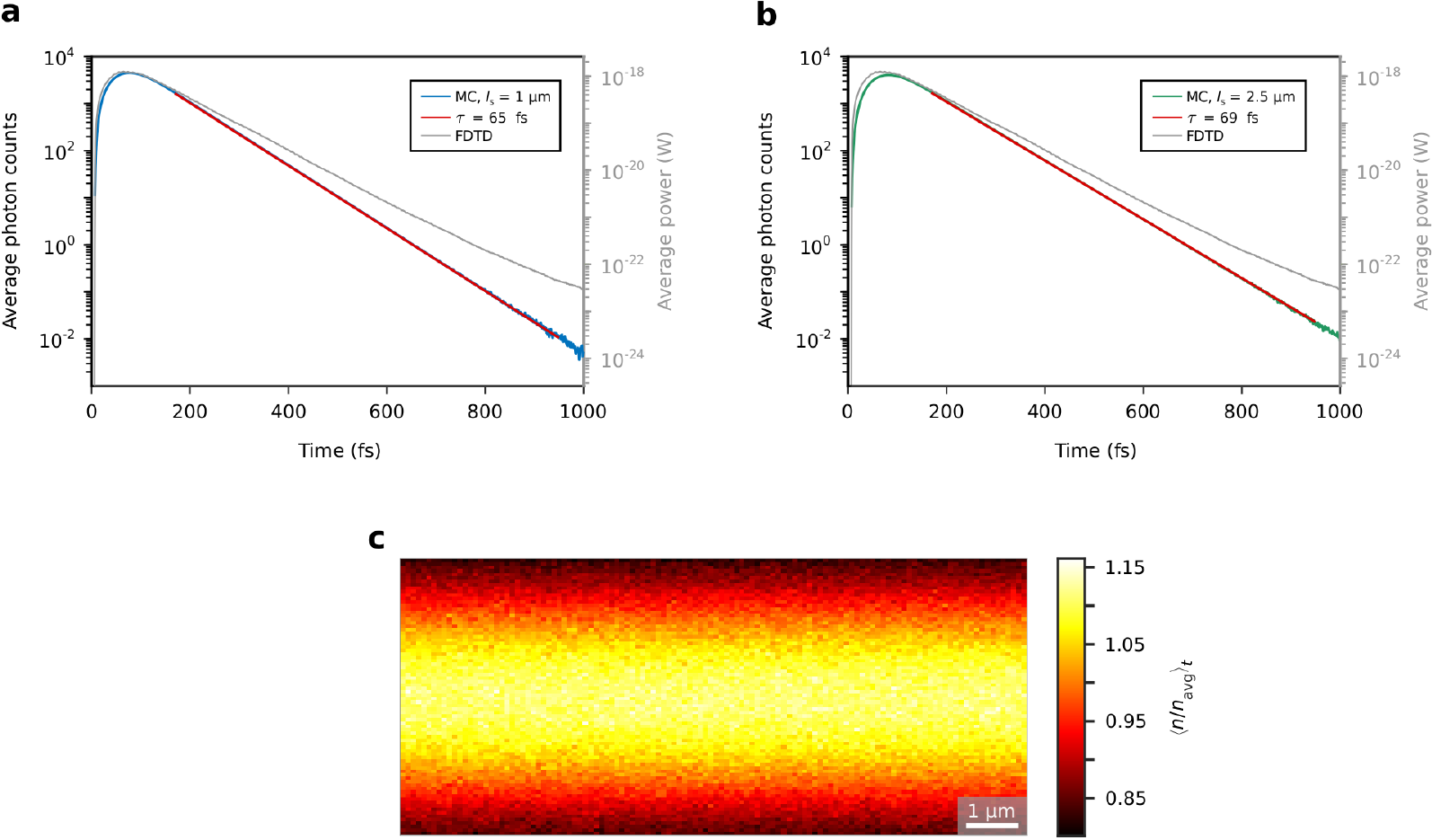}
\caption{\label{fig:comparison_ls}
\textbf{Evaluation of MC simulation parameters.}
\textbf{a}, Average photon counts in the monitor plane obtained by a MC simulation with a scattering mean free path $l_{\text{s}}$ of 1 µm. The mono-exponentially fitted tail (red line) reveals a decay constant of 65 fs. Additionally the average power calculated by a FDTD simulation is shown using the second ordinate. Both ordinates span the same order of magnitudes. 
\textbf{b}, same as \textbf{a}, but a scattering mean free path $l_{\text{s}}$ of 2.5 µm is used in the MC simulation and the fit (red line) exhibits a decay constant of 69 fs. 
\textbf{c}, Time averaged photon counts enhancement recorded in the monitor plane (span from 170 fs to 650 fs).}
\end{center}
\end{figure}

Due to the relationship $l_{\text{t}} = l_{\text{s}}/(1-g)$ between the transport mean free path $l_{\text{t}}$, the scattering mean free path $l_{\text{s}}$ and the anisotropy factor $g$ the scattering mean free path can be chosen arbitrary in the interval \linebreak (0, $l_{\text{t}}$] as long as $g$ is chosen adequately to match $l_{\text{t}}$. Since the isotropic diffusion equation does not explicitly depends on $l_{\text{s}}$ and $g$, respectively, the MC simulations indeed reveal independently of the applied $l_{\text{s}}$ the correct reflectance as long as the correct $l_{\text{t}}$ is used. A decent value for $l_{\text{s}}$ can be obtained comparing the results of the MC and FDTD simulations. As shown exemplarily in Fig. \ref{fig:comparison_ls}a and \ref{fig:comparison_ls}b a small $l_{\text{s}}$  of 1 µm is suitable to describe the FDTD results for short times correctly while for a larger $l_{\text{s}}$ of 2.5 µm the top of the curves deviate. However, in both cases the MC simulations fail to describe the tail correctly since the MC curves possess a mono-exponential decay in contrast to the multi-exponential decay found in FDTD simulations.\\
Analogous to the local power enhancement calculated using the FDTD results (Fig. \ref{fig:comparison_ls}d), for the MC simulations the photon counts are normalised to the average photon counts in the monitor plane for every time step. Averaging the time steps over a span ranging from 170 fs to 650 fs yields the result displayed in Fig. \ref{fig:comparison_ls}c. In contrast to FDTD (cf. Fig. \ref{fig:fig2}f, inset and Fig. \ref{fig:Details_FDTD}d, top), the MC simulation do not reveal any hotspots but an almost constant photon count enhancement in lateral direction. Due to the possibility that photons leave the slab at the upper and lower boundaries there is an axial profile possessing a slight enhancement in the middle of the slab. Nevertheless, this enhancement is only 15\% above the average not being a significant enhancement as for the hotspots in the FDTD simulations.

\renewcommand\refname{Supplementary References}